\documentclass{aa}  

\usepackage{graphicx}
\usepackage{txfonts}
\usepackage{hyperref}
\begin{document}

   \title{On the origin of mid-infrared colors in $\gamma$-ray blazars}
   \titlerunning{On the origin of mid-infrared colors in $\gamma$-ray blazars}

   \author{Raniere de Menezes$^{1,2}$\thanks{E-mail: raniere.m.menezes@gmail.com}, Raffaele D'Abrusco$^3$, Francesco Massaro$^{1,2,4,5}$ }
   \authorrunning{R. de Menezes et al.}

   \institute{INFN -- Istituto Nazionale di Fisica Nucleare, Sezione di Torino, via Pietro Giuria 1, I-10125 Turin, Italy
   \and
   Dipartimento di Fisica, Universit\`a degli Studi di Torino, via Pietro Giuria 1, I-10125 Torino, Italy
   \and
   Center for Astrophysics | Harvard \& Smithsonian, 60 Garden Street, Cambridge, MA 20138, USA
   \and
   INAF-Osservatorio Astrofisico di Torino, via Osservatorio 20, 10025 Pino Torinese, Italy
   \and
   Consorzio Interuniversitario per la Fisica Spaziale (CIFS), via Pietro Giuria 1, 10125 Torino, Italy
             }

   \date{Received October XX, 2024; accepted XXX}

 
  \abstract
   {The combination between non-thermal and thermal emission in $\gamma$-ray blazars pushes them to a specific region of the mid-infrared three-dimensional color diagram, the so-called \textit{blazar locus}, built based on observations performed with the Wide-field Infrared Survey Explorer. The selection of blazar candidates based on these mid-infrared colors has been extensively used in the past decade in the hunt for the counterparts of unassociated $\gamma$-ray sources observed with the Fermi Large Area Telescope and in the search for new blazars in optical spectroscopic campaigns.}
   {In this work, we provide a theoretical description of the origin of the \textit{blazar locus} and show how we can reasonably reproduce it with a model consisting of only three spectral components: a log-parabola accounting for the non-thermal emission, and an elliptical host and dust torus accounting for the thermal emission.}
   {We simulate spectral energy distributions (SEDs) for blazars, starting with a pure log-parabola model and then increasing its complexity by adding a template elliptical galaxy and dust torus. From these simulations, we compute the mid-infrared magnitudes and corresponding colors to create our own version of the \textit{blazar locus}.}
   {Our modeling allows for the selection of spectral parameters that better characterize the mid-infrared emission of $\gamma$-ray blazars, such as the log-parabola curvature ($\beta < 0.04$ for 50\% of our sample) and an average spectral peak around $E_p \approx 1.5 \times 10^{-13}$ erg. We also find that the log-parabola is the main spectral component behind the observed mid-infrared blazar colors, although additional components such as a host galaxy and a dust torus are crucial to obtain a precise reconstruction of the \textit{blazar locus}.}
   {}


   \maketitle
%

\section{Introduction}
\label{sec:intro}

Blazars are one of the most elusive types of active galactic nuclei (AGNs) and their multiwavelength emission, mainly non-thermal, arises from the acceleration of charged particles in a relativistic jet closely aligned with the line of sight \citep{blandford1979relativistic}. They are divided into two main classes based on their optical spectra \citep{massaro2009romabzcat}, i.e., the BL Lacs, which present only weak emission or absorption lines (equivalent widths $< 5$~\AA), or even a continuum-dominated spectrum completely depleted of lines \citep{stickel1991complete,landoni2014spectroscopy}; and the flat spectrum radio quasars (FSRQs), which have broad emission lines, a dominant blue continuum, and a flat radio spectrum \citep[spectral index $\alpha<0.5$ in the $1\sim5$ GHz range;][]{chen2009_FSRQs,ghisellini2011transition}. Although relatively rare among AGNs, blazars are the dominant population in the $\gamma$-ray sky, accounting for more than 50\% of all sources observed so far with the \textit{Fermi} Large Area Telescope \citep[LAT;][]{abdollahi2022_4FGL_DR3}.

The non-thermal emission of blazars produces a characteristic spectral energy distribution (SED) that exhibits two broad bumps: one at low energies, originating from the synchrotron emission of relativistic particles accelerated in the blazar jet and peaking somewhere between the radio and soft X-rays; and a second one peaking at $\gamma$-rays typically associated with the inverse Compton scattering of local synchrotron or external thermal photons by the relativistic leptons in the blazar jet \citep{fossati1998unifying,ghisellini1998theoretical,abdo2010_Fermi_blazar_SEDs}, although a hadronic interpretation is also possible \citep{bottcher2013leptonic,demenezes2020llagn}. This particular SED shape, dominated by non-thermal emission, positions $\gamma$-ray blazars in a distinct region of the mid-infrared three-dimensional color space defined by the Wide-field Infrared Survey Explorer \citep[WISE;][]{wright2010WISE_telescope} filters, the so-called \textit{blazar locus} \citep{massaro2011identification,massaro2012wise_UGS,dAbrusco2014wibrals}.

These mid-infrared properties of blazars have then been used as a diagnostic tool for the characterization of AGNs \citep{stern2012mid,assef2013mid,yan2013characterizing,mateos2013uncovering} and the identification/association of uncertain and unknown low-energy counterparts of \textit{Fermi}-LAT $\gamma$-ray sources \citep{massaro2012wise_UGS, massaro2015refining, massaro2016infrared, abdollahi2020_4FGL, deMenezes2020physical}, many of which were later confirmed to be blazars via optical spectroscopy \citep[e.g.,][]{paggi2013unveiling,ricci2015optical,crespo2016opticalVI, marchesini2019optical, deMenezes2020optical_camp_X, pena2020optical, pena2021optical,garcia2023optical_camp}. In this context, the WISE Blazar-Like Radio-Loud Sources catalog \citep[WIBRaLS;][]{dAbrusco2014wibrals,dAbrusco2019wibrals2_KDEBLLACS,deMenezes2019characterization} emerges as one of the most successful catalogs of $\gamma$-ray blazar candidates, which was designed by selecting radio-loud sources detected in all four WISE bands (nominally at 3.4, 4.6, 12, and 22 $\mu$m) and presenting mid-infrared colors similar to those of \textit{Fermi}-LAT blazars (i.e. within the \textit{blazar locus}).

In this work, we model the SED of sources located within the \textit{blazar locus}, showing why the blazars occupy a specific region of the three-dimensional mid-infrared color space, well separated from other astrophysical sources dominated by thermal radiation, and identifying the ranges of non-thermal parameters that characterize these blazars. We furthermore discuss why this \textit{locus} cannot be reasonably reproduced by sources with pure power-law spectra in the mid-infrared.

Throughout this work, the WISE bands are indicated as W1, W1, W3, and W4, corresponding respectively to the nominal wavelengths centered at 3.4, 4.6, 12, and 22 $\mu$m. The present paper is organized as follows. In \S \ref{sec:blazar_locus} we detail how the \textit{blazar locus} was built based on a clean selection of blazars. In \S \ref{sec:locus_modeling} we describe how we model the \textit{locus} based on a combination of non-thermal emission from the jet, thermal emission from the host galaxy, and thermal emission from a dust torus. We show the results of our modeling in \S \ref{sec:results} and present a discussion and conclusions in \S \ref{sec:discussion_and_conclusions}. In this work we assume a flat Universe with $h = 0.70$, $\Omega_m = 0.30$, and $\Omega_{\Lambda} = 0.70$, where the Hubble constant is $H_0 = 100~h$ km s$^{-1}$ Mpc$^{-1}$ \citep{tegmark2004cosmological}. The WISE magnitudes adopted here are in the Vega system and are not corrected for the Galactic extinction since such a correction only affects the W1 band for sources lying close to the Galactic plane, and it ranges between 2\% and 5\% of a magnitude \citep{dAbrusco2014wibrals}, thus not significantly affecting the results.


\section{The blazar locus}
\label{sec:blazar_locus}

In the two-dimensional $W1-W2 \times W2-W3$ color-color diagram for WISE sources, the blazars, which are dominated by non-thermal emission, occupy a very specific region (the so-called \textit{blazar strip}), well separated from other sources that are dominated by thermal radiation, as shown by \cite{massaro2011identification,massaro2012colours,massaro2012wise_UGS}. In following works \citep{dAbrusco2012infrared,dAbrusco2014wibrals}, a refined model of this region was built including a third axis (i.e. $W3-W4$) in the mid-infrared WISE color diagram, the so-called \textit{blazar locus}.

The \textit{blazar locus} adopted here is based on the original sample of blazars described in \cite{dAbrusco2019wibrals2_KDEBLLACS}. This sample consists of the blazars listed in the \textit{Fermi}-LAT third source catalog \citep[3FGL;][]{acero2015_3FGL} that have a counterpart in the Roma-BZCat catalog \citep{massaro2015romabzcat5th} and are detected in all four WISE filters. The counterpart in Roma-BZCat guarantees that the blazar nature of the source was carefully verified by inspection of its multi-wavelength emission, while the counterpart in 3FGL guarantees that these blazars are $\gamma$-ray emitters. We also verified that using the updated version of the Fermi-LAT catalog (i.e. 4FGL) increases the number of locus sources by less than 5\% and does not appreciably change the values of the best-fit parameters of the \textit{locus} model adopted here and published by \cite{dAbrusco2019wibrals2_KDEBLLACS}. The final sample used to define the WISE \textit{blazar locus} consists of 901 $\gamma$-ray-emitting blazars, split into 497 BL Lacs and 404 FSRQs.

\begin{figure*}
    \centering
    \includegraphics[width=\linewidth]{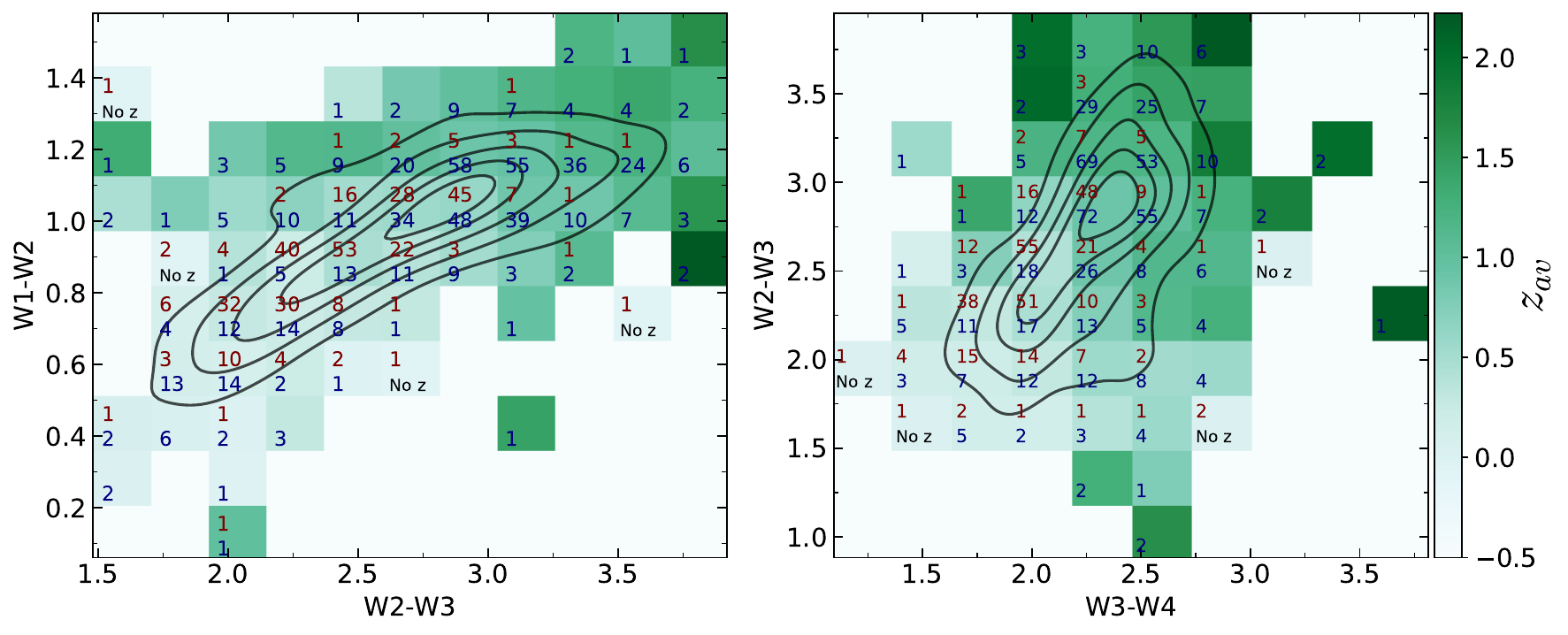}
    \caption{The distribution of average redshift (green tiles) for the blazars used to create the \textit{locus}. We see that the regions dominated by FSRQs (i.e. top-right corners of both panels) have higher average redshifts ($z_{av}$). The dark-blue numbers within each tile represent the total number of sources used to compute the average redshift, while the brown numbers represent the number of blazars with unknown redshift. Tiles with no redshift available are set to 0 and tagged with the label ``No z'', while the background tiles are set to -0.5. The black contours represent the 90\% containment projections of the \textit{blazar locus} in both mid-infrared color-color planes.}
    \label{fig:redshift_distribution_planes}
\end{figure*}

The intrinsic distribution of blazar redshifts varies between the two spectral classes, and the observational incompleteness of our sample affects these classes differently, leading to a reshaping of the \textit{locus} driven by the redder colors of high-redshift sources. In Figure \ref{fig:redshift_distribution_planes} we show how different regions of the \textit{blazar locus} (defined by the 90\% containment black contours) present different average redshifts, with the top-right corner of both panels being dominated by higher-redshift sources \citep[all redshifts were collected from][]{massaro2015romabzcat5th}. A realistic model for the \textit{locus} has to consider this fundamental feature (see \S \ref{subsec:adding_elliptical}). Each green tile in this figure represents the average redshift computed for the N blazars lying within the tile edges. Since we only have the redshifts for 559 blazars (out of 901), the average redshift per tile is computed only for those blazars with an available redshift (dark-blue numbers), while those sources with unknown redshifts (brown numbers) are not taken into account. For those cells containing more than 10 blazars with an available redshift, we also computed the redshift standard deviation, finding values in the range of $25\%$ (top-right corner of the \textit{locus}) to $\sim 50\%$ (bottom-left corner) of $z_{av}$.

\section{Locus modeling}
\label{sec:locus_modeling}

To model the lower-energy bump in the SED of blazars, we start with a semi-analytical model where a log-parabolic shape is used to describe the peak of the synchrotron emission \citep[see e.g.][for a qualitative and quantitative discussion on this topic]{landau1986active, massaro2004logpar1}. A log-parabolic spectrum is naturally obtained if the statistical acceleration of particles in the blazar jet has an energy-dependent probability that goes with $1/E^b$ (where $E$ is the particle's energy and $b$ is a positive constant), such that faster particles tend to escape from the acceleration site without being further accelerated \citep{massaro2004logpar1,massaro2004logpar2,massaro2006logpar3}. The log-parabolic curve is typically written as:

\begin{equation}
    F(E) = K(E/E_1)^{-\alpha-\beta\log(E/E_1)} ~~~ [\rm{cm}^{-2}~ \rm{s}^{-1}~ \rm{erg}^{-1}],
    \label{eq:logpar_classical}
\end{equation}
where $\alpha$ is the spectral index, $\beta$ is the curvature parameter, such that larger $\beta$ values imply on a stronger curvature, and $E_1$ is the pivot energy, which is typically set as a constant to avoid degeneration with the normalization constant $K$. In this work, however, we prefer to use the log-parabolic spectrum in the form \citep{massaro2004logpar2,tanihata2004evolution,tramacere2007signatures}:

\begin{equation}
    S(E) = S_p10^{-\beta\log^2(E/E_p)} ~~~ [\rm{erg}~ \rm{cm}^{-2}~ \rm{s}^{-1}],
    \label{eq:logpar_convenient}
\end{equation}
which conveniently gives us the SED peak value $S_p = E_p^2\,F(E_p)$, the energy at which the SED peak is located $E_p$, and the spectral curvature $\beta$. In the following subsections, we will first describe the simplified SED model consisting only of a log-parabola component and then switch to a more complex description of the SEDs by including an elliptical host galaxy and a dust torus. In the simplistic log-parabola assumption, the problem is reduced to basically finding the best-fit values of the parameters $S_p$, $E_p$ and $\beta$.

\subsection{Non-thermal emission}
\label{subsec:logpar_SED}

\begin{figure*}
    \centering
    \includegraphics[width=\linewidth]{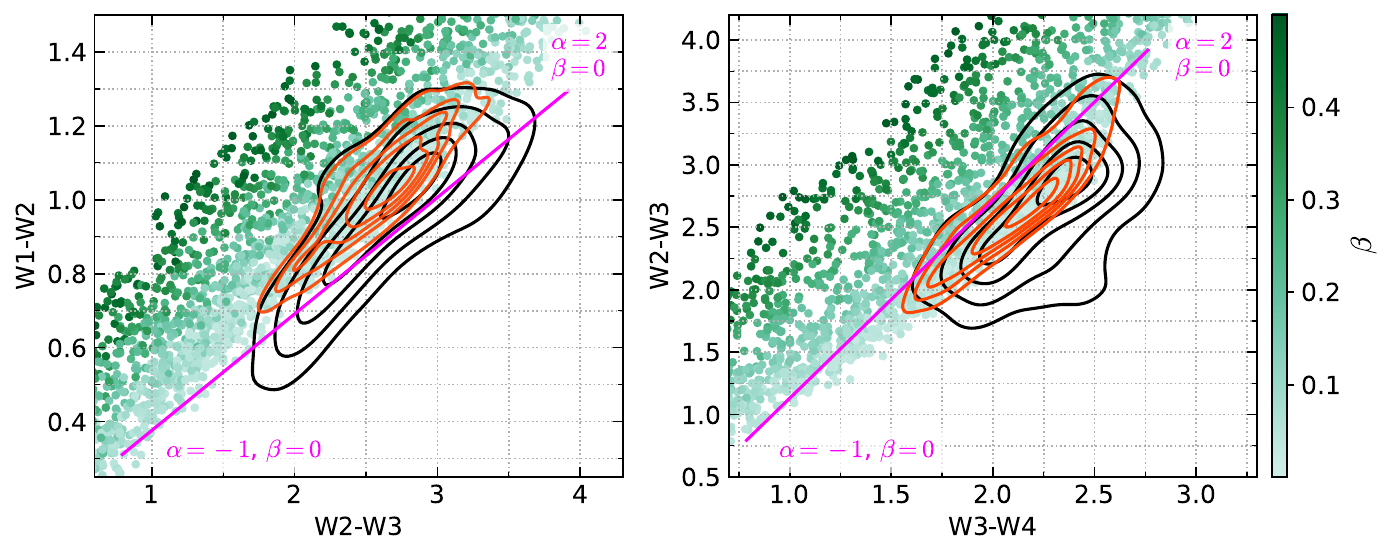}
    \caption{The mid-infrared color-color diagrams for the log-parabola model. We see that this simplified model already seems to suggest that the \textit{blazar locus} (represented by the three-dimensional color space delimited by the 90\%-containment black contours from both panels) is populated by blazars with weak spectral curvature (i.e. relatively small values of $\beta$), although pure power-law spectra (magenta lines), i.e. zero spectral curvature, also seem to be insufficient to create the observed distribution of sources in the \textit{locus}. The red contours represent the 90\% containment region for the simulated sources lying within the 90\% containment contours of the \textit{blazar locus}. The red and black isodensity contours are significantly different from each other, as detailed in the text. Here we plot only 5\,000 points for readability reasons.}
    \label{fig:logpar_color-color}
\end{figure*}

We use Eq. \ref{eq:logpar_convenient} to simulate SEDs with parameters randomly chosen in the log-space ranges $10^{-14} < S_p < 10^{-10} ~\rm{erg}~ \rm{cm}^{-2}~ \rm{s}^{-1}$ and $10^{-16} < E_p < 10^{-8}$ erg; and in the linear space range $0 < \beta < 0.5$. These intervals of values have been chosen so that $S_p$ can reach differential fluxes as high as the brightest known blazars \citep[e.g., see Figures 4 and 5 in][]{giommi2021_blazar_SEDs}, $E_p$ can be anywhere between the radio/far-infrared \citep[as can be the case for FSRQs, e.g.][]{giommi2012simplified_view_of_blazars, anjum2020origin} and soft X-rays \citep[as can be the case for high synchrotron peak blazars, e.g.][]{fossati1998unifying,bartoli2012long-term_Mrk501}, and $\beta$ assumes realistic synchrotron peak curvature values as discussed in \cite{chen2014curvature}. For each one of the models adopted herein, we simulate $10^5$ sources divided into 10 groups of $10^4$. The size of each group guarantees that the final number of sources within the locus 90\% containment contours is of the order $\lesssim 10^3$, which is of the same order as the original sample used to build the locus (i.e. 901 blazars).

We compute the sources' average fluxes in each one of the WISE bands according to:

\begin{equation}
    F_N = \frac{1}{\Delta E} \int_{E_{min,N}}^{E_{max,N}} \frac{S(E~|~\vec\theta~)}{E} dE,
    \label{eq:diff_fluxes}
\end{equation}
where $N$ corresponds to one of the 4 WISE bands, $S(E)$ is defined in Eq. \ref{eq:logpar_convenient}, $E_{max,N}$ and $E_{min,N}$ are the energy limits for each WISE band, and $\vec\theta$ represents the set of log-parabola parameters randomly selected from the ranges discussed above. 
The final values of $F_N$ are then converted to Jansky and their respective magnitudes in the Vega System are calculated according to \citep[see][for further details on WISE magnitudes]{wright2010WISE_telescope,jarrett2011spitzer}\footnote{A summary about WISE magnitudes can also be found here: \url{https://wise2.ipac.caltech.edu/docs/release/allsky/expsup/sec4_4h.html}}:

\begin{equation}
    W_N = -2.5\log_{10} \left( \frac{f_c F_{\nu,N}}{F_{\nu 0,N}} \right),
\end{equation}
where $W_N$ represents one of the four WISE magnitudes, $F_{\nu,N}$ is the target flux density in the $N$ band in Jy, $F_{\nu 0,N}$ is the zero magnitude flux density in the $N$ band derived for sources with power-law spectra $F_{\nu} \propto \nu^{-2}$, and $f_c$ is a correction factor dependent on the local power-law spectral index. The four values for $F_{\nu 0,N}$ can be found in Section 2.2 in \cite{wright2010WISE_telescope}. To incorporate observational fluctuations into our model (mainly driven by variability), we add a 5\% Gaussian noise to each value of $F_{\nu,N}$ before computing the magnitudes. The mid-infrared color-color diagrams for this simplified model can be found in Fig. \ref{fig:logpar_color-color}, where we immediately see that the \textit{blazar locus} is mainly populated by blazars with relatively low values of $\beta$, and that a simple log-parabolic SED cannot fully reproduce the range of mid-infrared colors in the \textit{blazar locus}, delimited here by the outermost 90\% containment isodensity black contour. The isodensity curves (90\% containment) for simulated sources lying within the \textit{locus} 90\% containment black contours are shown in red and are compared with the original distribution of sources in the \textit{blazar locus} via a two-dimensional Kolmogorov-Smirnov (KS) test \citep{fasano1987multidimensional}\footnote{Available as the Python module \texttt{ndtest} here: \url{https://github.com/syrte/ndtest?tab=readme-ov-file}}, giving average two-tailed p-values and KS statistics of $P = 10^{-21.46 \pm 2.48}$ and $D = 0.27 \pm 0.02$ (left panel), and $P = 10^{-20.31 \pm 0.87}$ and $D = 0.27 \pm 0.01$ (right panel), where the uncertainties are derived as the $\log_{10} P$ and $D$ standard deviations for the 10 groups described above. With such small p-values, this simplified model is rejected as the origin of the \textit{blazar locus}. Ideally, we want p-values larger than $10^{-6.24}$, implying that our model cannot be rejected as the origin of the \textit{locus} at the $5\sigma$ Gaussian-equivalent confidence level. At this stage of the modeling, we do not consider the redshifts of the targets, but we will do it in the following sections as the model becomes more complex.

In Fig. \ref{fig:logpar_color-color} we also show magenta straight lines corresponding to the mid-infrared colors of sources with a pure power-law spectrum (i.e., Eq. \ref{eq:logpar_classical} with $\beta = 0$), with spectral indices in the range $-1 \leq \alpha \leq 2$. Although small values of $\beta$ are overall favored (more details in \S \ref{sec:results}), the distribution of sources in the \textit{blazar locus} is not symmetrical around the $\beta = 0$ line, suggesting that sources with pure power-law spectra cannot fully populate the locus \citep[see][ for a discussion on this topic]{massaro2011identification}.

\subsection{Blazar host galaxy}
\label{subsec:adding_elliptical}

\begin{figure}
    \centering
    \includegraphics[scale=0.6]{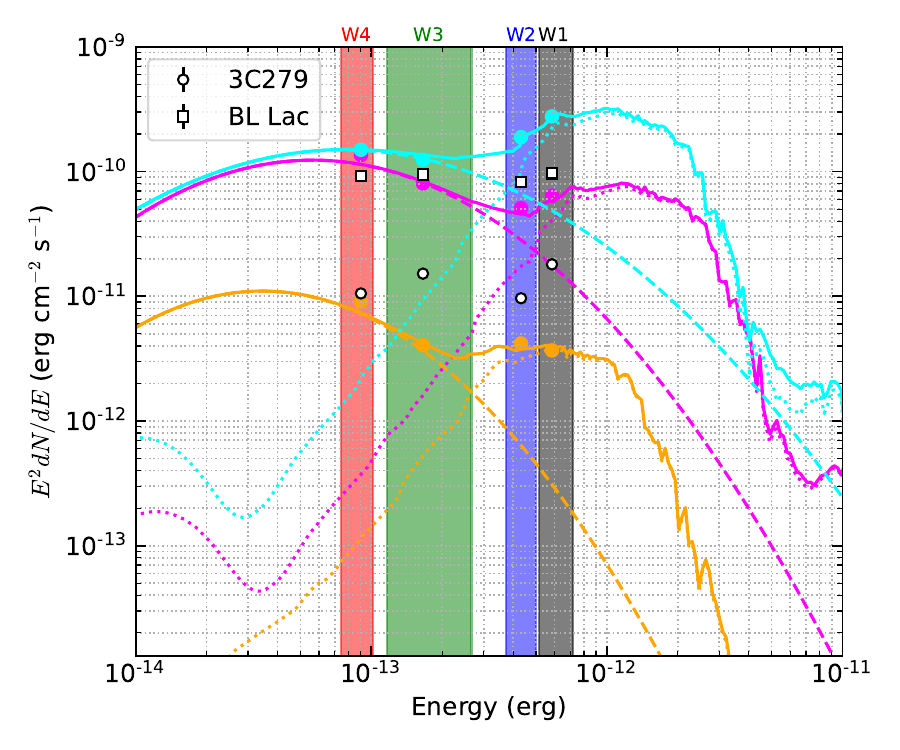}
    \caption{The SEDs of three randomly chosen blazars from our simulations, each one represented by a specific color. The final SED from which we compute the magnitudes consists of the sum (solid lines) of a galactic (dotted lines) and a log-parabola (dashed lines) component. In the WISE bands (vertical-colored stripes) where the log-parabola component is relatively weak, we see a significant contribution from the host galaxy. The final differential fluxes (see text for details), including the 5\% Gaussian noise, for each blazar in each band are shown as filled circles. For comparison, we also show the SED points measured with WISE for the blazars 3C279 and BL Lac, both of which are included in the \textit{locus} sample.}
    \label{fig:SEDs}
\end{figure}

BL Lacs and FSRQs are typically hosted by elliptical galaxies \citep{urry2000host_gal_of_bllacs,ODowd2002host_gals_radio-loud_AGN,olguin2016host}. At this stage of the modeling, we add an elliptical galaxy component to the log-parabolic SEDs described in \S \ref{subsec:logpar_SED}. The galactic templates are collected from the SWIRE Template Library \citep{polletta2007spectral}\footnote{The templates are available online at the following URL: \url{http://www.iasf-milano.inaf.it/~polletta/templates/swire_templates.html}} and cover rest-frame wavelengths from 0.1 $\mu$m up to 1000 $\mu$m (i.e. from the far ultraviolet to the far infrared). We select three galactic templates (namely Ell2, Ell5, and Ell13, following the SWIRE nomenclature) and redshift them by a random value in the range $0.001 < z < 4.5$ with selection weights based on the two redshift distributions (i.e. one for BL Lacs and one for FSRQs) of the sample used to build the \textit{blazar locus}. We furthermore normalize each template such that their bolometric luminosities, $L^{bol}$, are in the range $10^{-5}L_{cD}^{max} < L^{bol} < L_{cD}^{max}$, where $L_{cD}^{max} \approx 10^{44}$ erg/s approximates the bolometric luminosity limit of cD galaxies, and calculate the differential fluxes by adding a component $S_{gal}(E)$ into the integrand of Eq. \ref{eq:diff_fluxes}. In Fig. \ref{fig:SEDs} we show three simulated SEDs randomly chosen from our sample and compare them with the SED data points of the blazars 3C279 and BL Lac (both of which are included in the \textit{locus} sample). It is clear how the host galaxy's thermal emission can significantly affect the mid-infrared colors for a given target, especially in the WISE bands W1 and W2 (represented by the black and blue vertical stripes, respectively); and how galaxies at higher redshifts will present substantially different mid-infrared colors. In this figure we also see that the computed SED points (filled circles) are consistent with the measured SED points for 3C279 (empty circles) and BL Lac (empty squares) in terms of shape and normalization.

\begin{figure*}
    \centering
    \includegraphics[width=\linewidth]{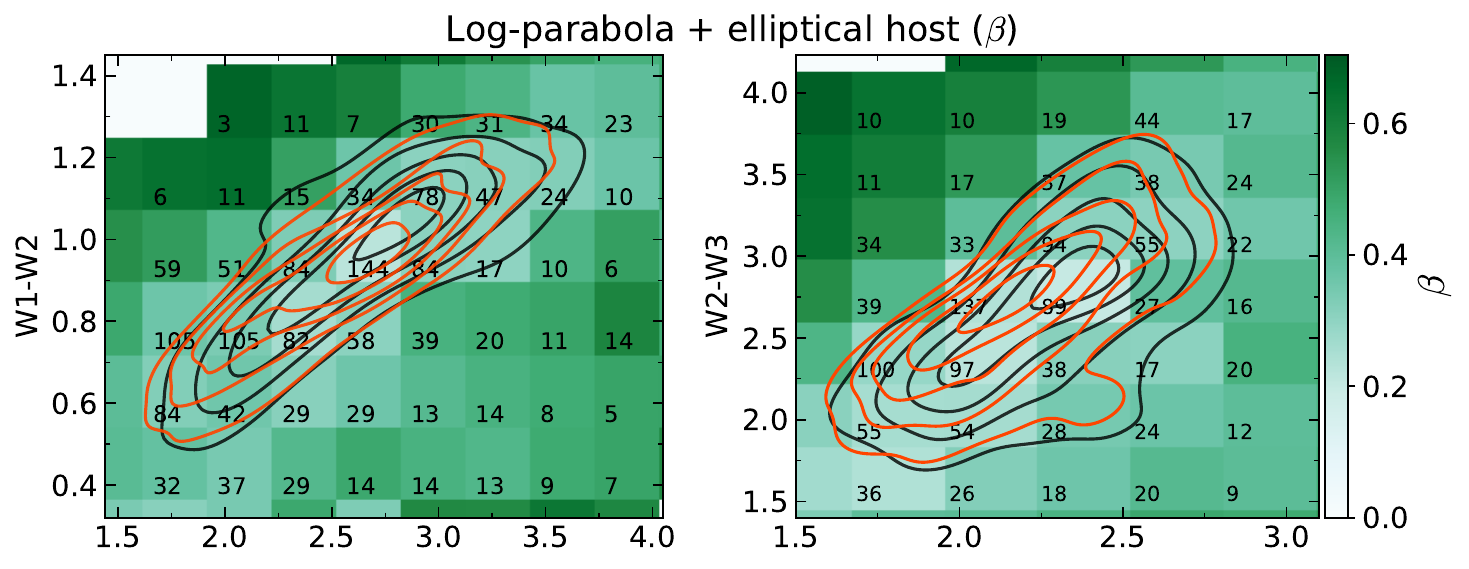}
    \includegraphics[width=\linewidth]{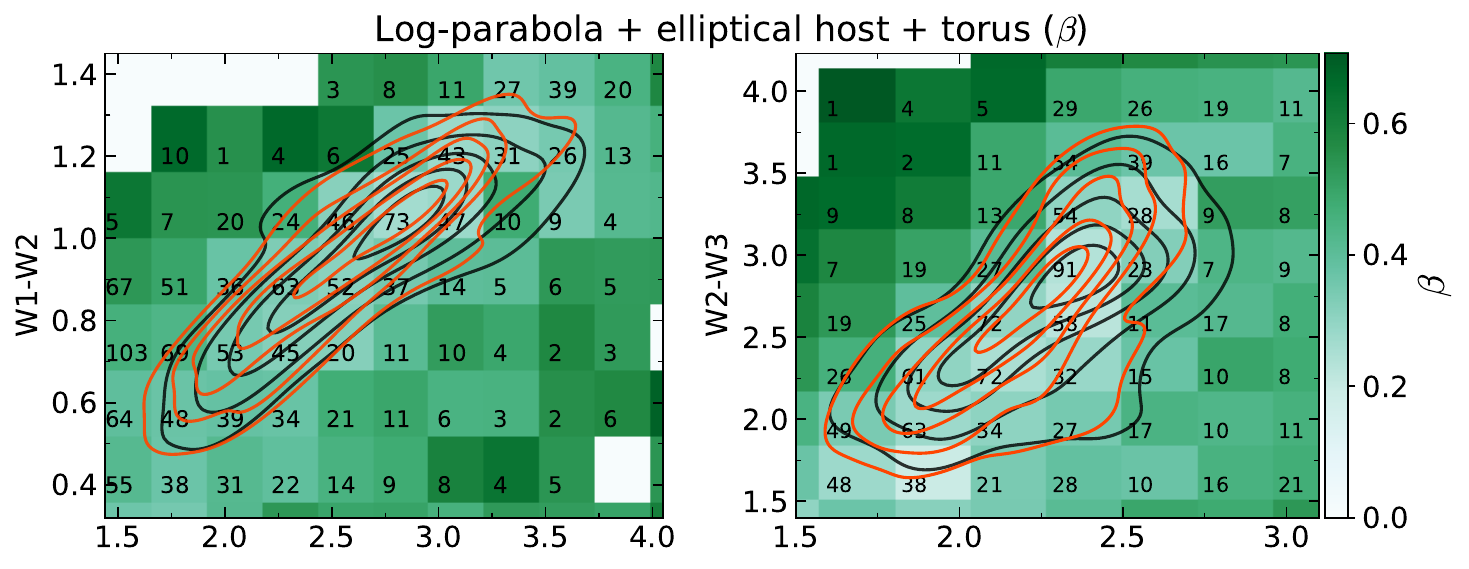}
    \includegraphics[width=\linewidth]{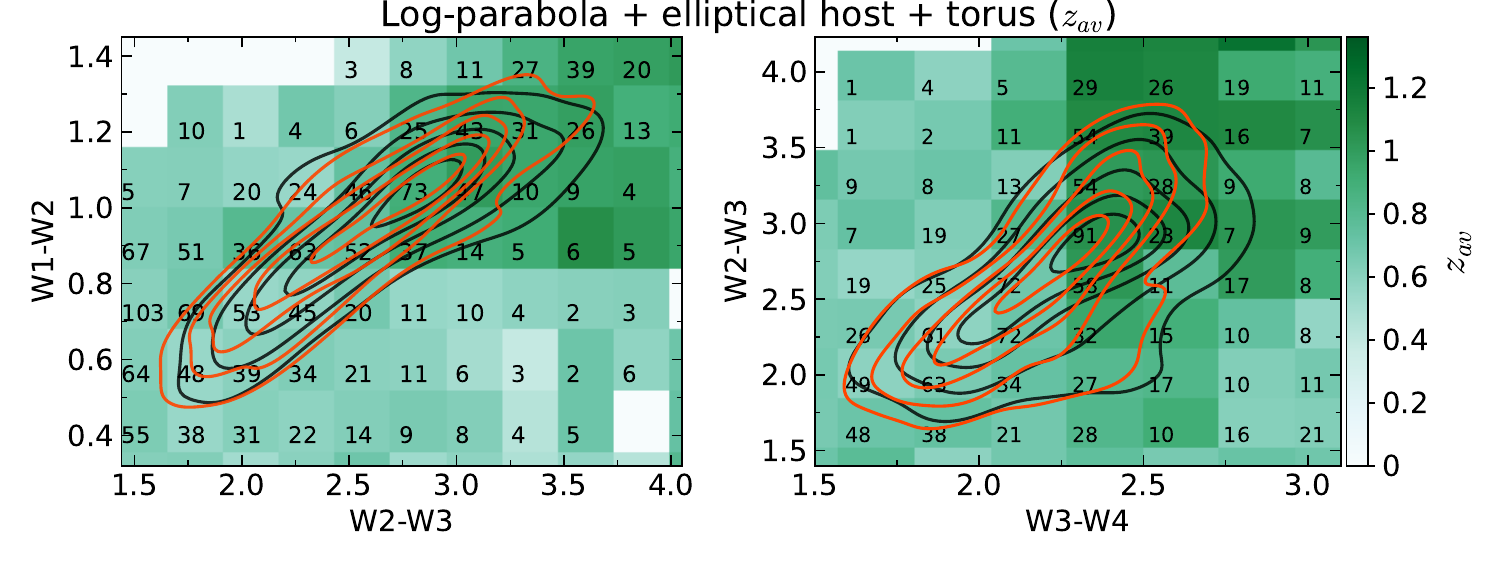}
    \caption{The mid-infrared color-color diagrams for models consisting of a log-parabola and a host elliptical galaxy components (top panels), and the same model with the addition of a dust torus for the FSRQs (middle panels). The overall distribution of simulated sources within the \textit{locus} (90\% containment red contours) for this later model is in much better agreement with the original distribution of sources in the \textit{blazar locus} (90\% containment black contours) if compared with the previous models. In the bottom panels, we show how the average blazar redshift is distributed in our simulations (for the same model as in the middle panels), which agrees with Fig. \ref{fig:redshift_distribution_planes}. All of these panels represent only one simulation (out of 10) with 10\,000 sources.}
    \label{fig:ellipticals_color-color}
\end{figure*}

In the top panels of Fig. \ref{fig:ellipticals_color-color} we show the mid-infrared color-color diagrams for this upgraded model. We see that the red contours (90\% isodensity curves for the simulated sources lying within the 90\% containment contour of the \textit{blazar locus}) already give us a much better representation of the \textit{blazar locus}, with two-tailed p-values and KS statistics of $P = 10^{-6.71 \pm 1.39}$ and $D = 0.15 \pm 0.01$ (left panel), and $P = 10^{-8.88 \pm 1.45}$ and $D = 0.18 \pm 0.01$ (right panel), which is again permeated by sources with relatively small $\beta$ values. This upgraded model is especially good at recovering the bottom section of both color-color \textit{locus} projections, i.e., the regions dominated by BL Lacs, although the relatively small p-values discussed above guarantee that something is still missing. In the next section, we implement our model by adding a torus component.

\subsection{Adding a dust torus}
\label{subsec:adding_torus}

We repeat the analysis performed in \S \ref{subsec:adding_elliptical} separating the sources into two groups based on their mid-infrared colors, the first one corresponding to FSRQs and containing 80\% of the sources with $W1-W2 > 0.9$ and $W3-W4 > 2.2$, and the second one containing the remaining sources consisting mainly of BL Lacs. Selecting different color cuts in the ranges $ 0.8 < W1-W2 < 1.0$ and $2.1 < W3-W4 < 2.4$ does not significantly affect our results, as the fraction of BL Lacs returned by using these limits is always in the range $20\% \sim 25\%$.

We then add a dust torus component (again from the SWIRE archive) only for the first group, since FSRQs frequently show signs of thermal emission from a dust torus \citep{malmrose2011emission}, while there is no observational evidence for tori in BL Lacs \citep{plotkin2012lack}. The mid-infrared color distributions delivered by this model show a significant improvement in the $W1-W2 \times W2-W3$ projection of the \textit{blazar locus} (see the middle panels of Fig. \ref{fig:ellipticals_color-color}), with a two-tailed p-value and KS statistics of $P = 10^{-5.05 \pm 1.19}$ and $D = 0.13 \pm 0.01$, indicating that our model cannot be rejected as the origin of this \textit{locus} projection at the $4\sigma$ confidence level, given that the $4\sigma$ threshold for a two-tailed p-value is given by $P_{4\sigma} = 10^{-4.19}$, which is consistent with our results within the errors. For the $W2-W3 \times W3-W4$ projection, on the other hand, the results are slightly better than in the previous model, with $P = 10^{-7.19 \pm 1.40}$ and $D = 0.16 \pm 0.01$. In any case, this p-value guarantees that our model cannot be rejected at the $5\sigma$ confidence level (i.e. $P_{5\sigma} = 10^{-6.24}$). In both color-color projections, we see that the \textit{locus} is still filled with weak spectral curvature blazars (see \S \ref{sec:results} for details).

In the bottom panels of Fig. \ref{fig:ellipticals_color-color} we show the average redshift distribution for the model including the dust torus, which is quite similar to the original \textit{locus} distribution shown in Fig. \ref{fig:redshift_distribution_planes}, indicating that our simulations can reasonably reproduce this feature. With this model in hand, we now want to understand which parameters from Eq. \ref{eq:logpar_convenient} best describe the \textit{blazar locus}. We then select all the simulated sources lying within the 90\% containment contours of both \textit{locus} projections and investigate their parameter distributions, as shown in \S \ref{sec:results}. With this three-component model, we arrive as far as the observational constraints allow us and reach a reasonable description (i.e. rejection level $\lesssim 5\sigma$) of the \textit{blazar locus}.


\section{Results}
\label{sec:results}

\begin{figure*}
    \centering
    \includegraphics[width=\linewidth]{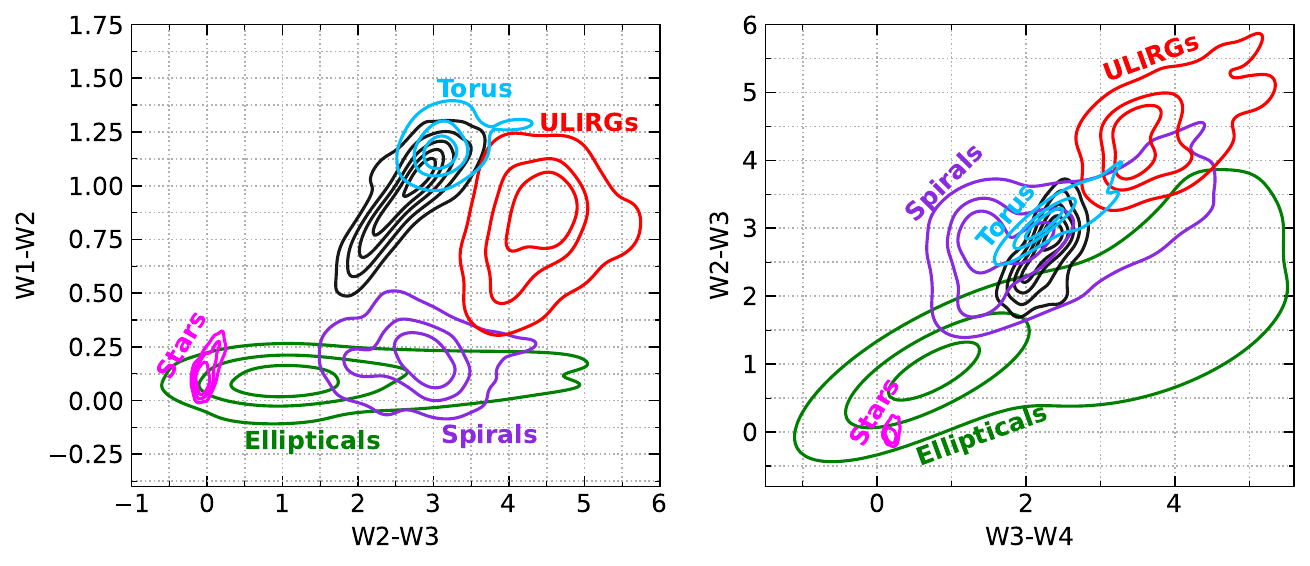}
    \caption{The mid-infrared color-color diagram 90\% containment regions dominated by elliptical galaxies (green), spiral galaxies (violet), ULIRGs (red), dust torus (cyan), and stars (magenta). We see that none of these components, by themselves, can completely fill the \textit{locus}, endorsing how important the log-parabola component (see Fig. \ref{fig:logpar_color-color}) is for our model.}
    \label{fig:color-color_regions}
\end{figure*}

Although a log-parabolic component alone is not enough to fully describe the \textit{blazar locus}, as shown in Fig. \ref{fig:logpar_color-color}, it is the most fundamental ingredient for all models tested here. In Fig. \ref{fig:color-color_regions} we see that the colors of elliptical galaxies and tori, by themselves, cannot fulfill the \textit{locus}, i.e., they need a substantial contribution from the log-parabola component. This component is so important that measuring its parameters (see Eq. \ref{eq:logpar_convenient}) can already give us a good characterization of the blazars found in the \textit{locus}. In Fig. \ref{fig:color-color_regions} we also highlight the zones (90\% containment contours) in the mid-infrared color space corresponding to the colors of spiral galaxies, ultra-luminous infrared galaxies (ULIRGs; where both templates are collected from the SWIRE archive), and nearly black-body spectra representing stars with surface temperatures in the range $2500~\rm{K} < T_{surf} < 7500$ K. We see that spiral galaxies represent a possible source of contamination in the $W2-W3 \times W3-W4$ \textit{torus} projection, however, this problem is attenuated by the W1-W2 selection performed for the 3D \textit{locus}.

\begin{figure*}
    \centering
    \includegraphics[width=\linewidth]{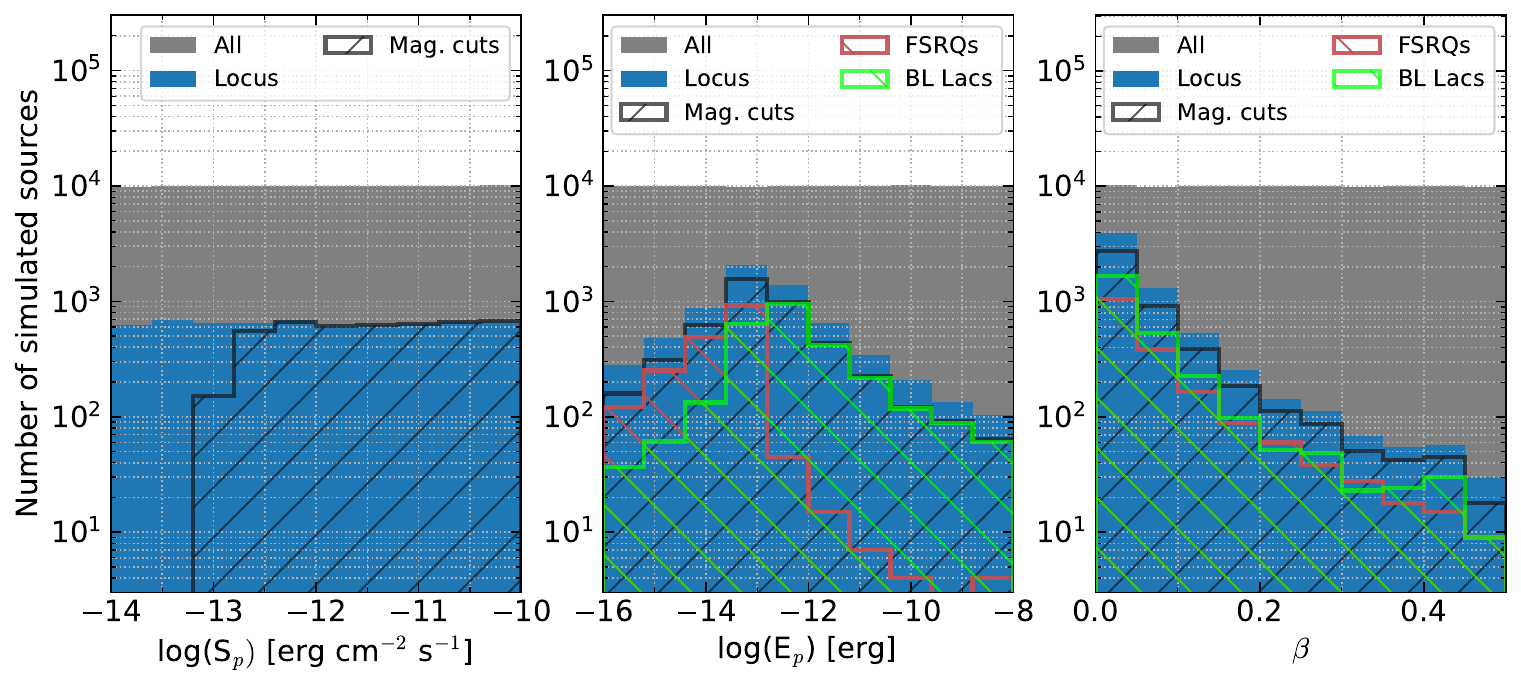}
    \caption{Distribution of log-parabola parameters (from Eq. \ref{eq:logpar_convenient}) for 100\,000 sources simulated with a model consisting of a log-parabola, an elliptical galaxy host, and a dust torus (grey histogram). We see that, overall, the sources within the \textit{blazar locus} (blue histogram) present weak spectral curvature ($\beta < 0.04$ for 50\% of the sample) and concentrate around $E_p \approx 1.5 \times 10^{-13}$ erg. The black hatched histograms represent the \textit{blazar locus} after we apply the WISE magnitude cuts and can be divided into two components, one with FSRQs (red histograms), and the other with BL Lacs (green histograms).}
    \label{fig:LP_parameters}
\end{figure*}

In Fig. \ref{fig:LP_parameters} we show the distribution of log-parabola parameters for 100\,000 sources simulated with the model described in \S \ref{subsec:adding_torus}. We immediately see that the simulated sources lying within the \textit{blazar locus} (blue histogram) present relatively small curvatures, where 50\% of the sample has $\beta < 0.04$ and 90\% has $\beta < 0.18$. Furthermore, the average energy peak is centered at $E_p \approx 1.5 \times 10^{-13}$ erg (i.e. within WISE band W3), with nearly 50\% of the blazars having $10^{-13.7} < E_p < 10^{-12.3}$ erg and nearly 90\% having $10^{-15} < E_p < 10^{-11}$ erg.

This favored value of $E_p$ tells us that, if the log-parabola spectral component peaks near the center (in log scale) of the range covered by the WISE filters, then it is basically guaranteed that WISE will detect it as a non-thermal source. The black hatched histograms in Fig. \ref{fig:LP_parameters} represent the distribution of log-parabola parameters in the \textit{locus} after cutting from our sample those sources with magnitudes below the WISE sensitivity limits, rounded to $W1 \lesssim 17.5$, $W2 \lesssim 16.5$, $W3 \lesssim 13.0$, and $W4 \lesssim 10.0$. As expected, the only parameter that is modified by the magnitude cuts is the SED energy flux peak $S_p$.

It is evident from the middle panel of Fig. \ref{fig:LP_parameters} that BL Lacs (green histogram) and FSRQs (red histogram) have different distributions of $E_p$, with the former peaking at $E_p \approx 10^{-12.5}$ erg and the latter at $E_p \approx 10^{-13.2}$ erg, i.e. just outside the energy range covered by the four WISE bands. The $\beta$ distributions, on the other hand, have a very similar shape, with BL Lacs slightly allowing for stronger curvatures.

\begin{figure}
    \centering
    \includegraphics[scale=0.6]{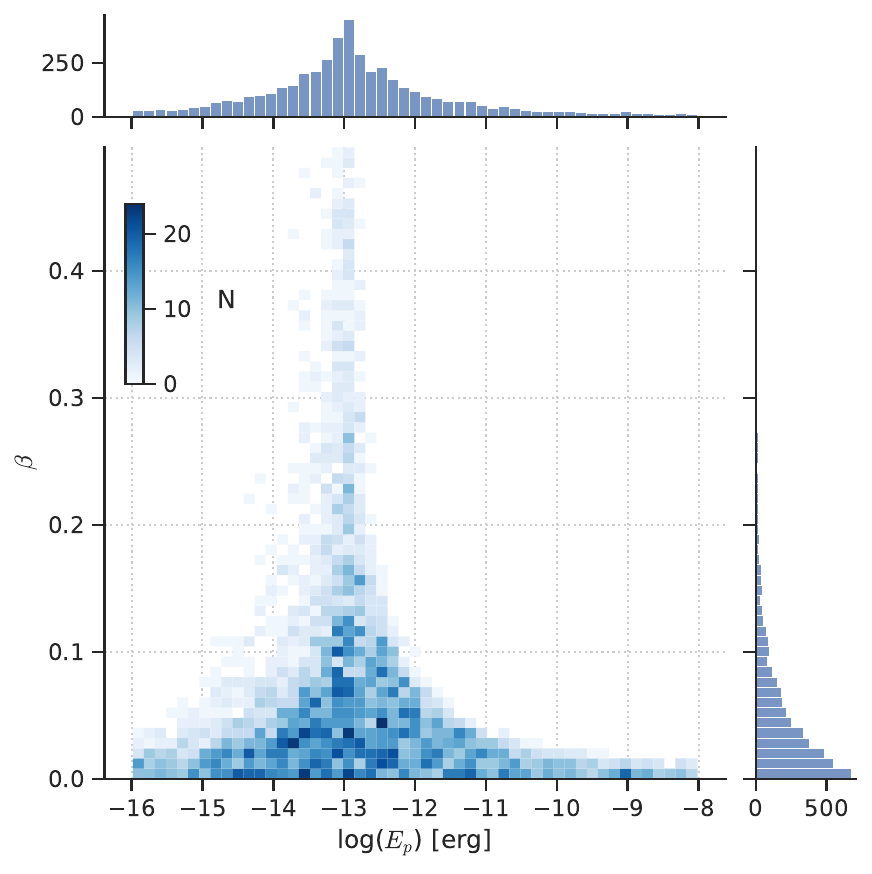}
    \caption{The distribution of the curvature parameter, $\beta$, in terms of the position of the log-parabola energy peak, $E_p$, for the model consisting of a log-parabola, a host elliptical galaxy and a dust torus (see \S \ref{subsec:adding_torus}). The color bar represents the number, $N$, of simulated sources found in each bin. We see that sources peaking in the range $10^{-14} \lesssim E_p \lesssim 10^{-12}$ erg allow for a wider range of $\beta$.}
    \label{fig:beta_x_Ep}
\end{figure}

We found no significant correlation between the parameters $\beta \times E_p$, for which we found a Pearson correlation coefficient $P_c = -0.097$ (or $P_c = -0.089$ if we use $\log_{10} E_p$); neither for the parameters $\log_{10} L_p \times \beta$ or $\log_{10} E_p \times \log_{10} L_p$, where $L_p \equiv S_p 4\pi d_L^2$ is the peak differential luminosity (i.e. $d_L$ is the luminosity distance) and the Pearson correlation coefficients are $P_c = 0.013$ and $P_c = 0.033$, respectively. We further notice that log-parabolas peaking in the range $10^{-14} 
 \lesssim E_p \lesssim 10^{-12}$ erg allow a wider range of $\beta$, as shown in Fig. \ref{fig:beta_x_Ep}. 

\section{Discussion and conclusions}
\label{sec:discussion_and_conclusions}

The full extent in the WISE color space occupied by the \textit{blazar locus} can be reasonably reproduced by a model only if it contains a combination of thermal and non-thermal emission components. The observations of elliptical galaxies as the hosts of BL Lacs objects \citep{falomo1996host,Kotilainen1998host,urry1999hubble,urry2000host_gal_of_bllacs} strongly suggest that the thermal emission from these hosts is a fundamental piece of a spectral model that correctly reproduces the mid-infrared colors of blazars \citep[in some rare cases, however, we can find disk galaxies as the hosts of BL Lacs;][]{abraham1991optical,urry2000host_gal_of_bllacs}. We also have observational evidence for the presence of dust tori in FSRQs \citep{malmrose2011emission} and, given that these are particularly bright in the mid-infrared, this component also seems to be necessary for the correct reconstruction of the \textit{blazar locus}. Based on these observations, in this work we focused our efforts on describing the \textit{blazar locus} with a model consisting of a log-parabola, a host elliptical galaxy, and a dust torus. This model seems to reasonably reconstruct the \textit{blazar locus} and cannot be rejected with a confidence greater than $4\sim 5 \sigma$ according to the KS test described in \S \ref{subsec:logpar_SED}. We reach several conclusions based on this modeling, as listed below:

\begin{itemize}
    \item The log-parabola is the main spectral component for blazars, although it cannot fully reproduce the colors of the locus by itself.
    \item An elliptical galaxy and a dust torus components \citep[whose presence is supported by observations, e.g.][]{urry2000host_gal_of_bllacs,malmrose2011emission} are necessary to fully populate the area of the WISE 3D color space occupied by the \textit{blazar locus}.
    \item Assuming simulated SEDs including a log-parabola, an elliptical host and a dust torus components (see \S \ref{subsec:adding_torus}), sources matching the position of the \textit{locus} tend to have relatively weak spectral curvatures, i.e. $\beta < 0.04$ for 50\% of the sample, and a distribution of spectral peaks centered at $E_p \approx 1.5 \times 10^{-13}$ erg.
    \item The average log-parabola spectral peak, $\langle E_p \rangle$, for BL Lacs is located nearly one order of magnitude at higher energies than for FSRQs.
    \item For log-parabola spectral components peaking around $10^{-14}~ \rm{erg} \lesssim E_p \lesssim 10^{-12}$ erg, the spectral curvatures can reach at least $\beta \approx 0.5$.
    \item Spirals contaminate the projection of the \textit{locus} on the W2-W3 $\times$ W3-W4 plane but do not significantly affect the \textit{locus} since they are filtered out by the color $W1-W2$.
\end{itemize}

The results described in this work provide the first semi-analytical theoretical ground for several previous articles on the mid-infrared observational properties of $\gamma$-ray blazars \citep[e.g.:][]{massaro2011identification,massaro2012colours,dAbrusco2012infrared,dAbrusco2014wibrals,dAbrusco2019wibrals2_KDEBLLACS,deMenezes2019characterization} and help on the quest for the association of \textit{Fermi}-LAT unidentified $\gamma$-ray sources with their low-energy counterparts \citep{paggi2014optical,landoni2015optical,massaro2016extragalactic,pena2019optical,deMenezes2020physical}.

\begin{acknowledgements}
      The authors express their gratitude to the anonymous referee for the constructive feedback, which has helped enhance the quality of the manuscript. R.M. acknowledges support from the \textit{Universit\`a degli Studi di Torino} and \textit{Istituto Nazionale di Fisica Nucleare} (INFN), \textit{Sezione di Torino}, under the \textit{assegno di ricerca Sviluppo di sensori basati su SiPM per ricerca di sorgenti di fotoni di $E>100$ GeV (finanziamento MIUR Dipartimenti di Eccellenza 2018-2022 – per il Dipartimento di Fisica), DFI.2021.24}. R.D'A. is supported by NASA contract NAS8-03060 (Chandra X-ray Center). In this work we extensively used \texttt{TOPCAT} \citep{taylor2005topcat} and \texttt{astropy} \citep{astropy2013A&A...558A..33A,astropy2018AJ....156..123A} for preparation and manipulation of the data. This publication makes use of data products from the Wide-field Infrared Survey Explorer, which is a joint project of the University of California, Los Angeles, and the Jet Propulsion Laboratory/California Institute of Technology, funded by the National Aeronautics and Space Administration.
\end{acknowledgements}


\bibliographystyle{aa}
\bibliography{aanda} 

\end{document}